\documentclass[12pt]{article}
\usepackage{latexsym,amsmath,amssymb,amsbsy,graphicx}
\usepackage[cp1251]{inputenc}
\usepackage[T2A]{fontenc}
\usepackage[russian,english]{babel}
\usepackage[space]{cite} 

\begin{document}

{\bf Powerful solar flares in September 2017. Comparison with the
largest flares in cycle 24 }

\bigskip

\centerline {E.A. Bruevich, V.V. Bruevich}

\centerline {\it  Lomonosov Moscow State University, Sternberg Astronomical Institute,}
\centerline {\it Universitetsky pr., 13, Moscow 119992, Russia}\

\centerline {\it e-mail:  {red-field@yandex.ru} }\

{\bf Abstract.} Solar flare activity in cycle 24 is studied. Satellite observations of x-ray fluxes from GOES-15 and UV
emission lines from the SDO/EVE experiment are used. The most powerful flares of cycle 24 in classes
X9.3 and X8.2 in September 2017 are compared with powerful flares in classes M5-X6.9. The times at
which the fluxes in the 30.4 and 9.4 nm lines and in the 0.1-0.8 nm x-ray range begin to increase are
compared for 21 of the large flares. The total energies arriving at the earth from flares in the 30.4 and
9.4 nm lines and in the 0.1-0.9 nm x-ray range, $E_{30.4}$, $E_{9.4}$, and $E_{0.1-0.8}$, from 25 flares during 2011 and 2012
are calculated. It is shown that the calculated energies of the flares in the analyzed lines from SDO/EVE
and in the x-ray range from GOES-15 are closely interrelated.

\bigskip

\vskip12pt
\section{Introduction}
\vskip12pt

 The most powerful flares observed on the sun eject immense energy into the surrounding space, roughly a
fifth of the energy released by the sun per second (for comparison, this equals all the energy released by humanity
over a million years at current rates). Among the stars with activity of the solar type (e.g., UV Cet-type flare stars), the sun has relatively low flare activity [1,2].
The current solar activity cycle 24 is currently approaching the lowest number of solar spots and other global
indices. There have been about 800 large flares in cycle 24 over the entire cycle according to an x-ray classification
based on measurements by the GOES series of satellites (the classes >M1 correspond to flares with amplitudes
exceeding $1 \cdot 10^{-5} W/m^2$ in the 0.1-0.8 nm range), while there were about twice this number in the previous, more
powerful cycles 22 and 23. The most powerful flare in cycle 24 up to September 2017 is considered to be in x-ray
class X6.9 and took place on August 9, 2011. It does not have entirely standard characteristics for such a large flare;
in particular, in terms of the amount of energy incident on the earth in the 0.1-0.8 nm x-ray range, it was not even
among the first ten. The flare of August 9, 2011 was the most substantial in cycle 24 through September 2017. The
features of this class X6.9 flare are discussed in detail in Refs. 3 and 4.

\begin{figure}[tbh!]
\centerline{
\includegraphics[width=120mm]{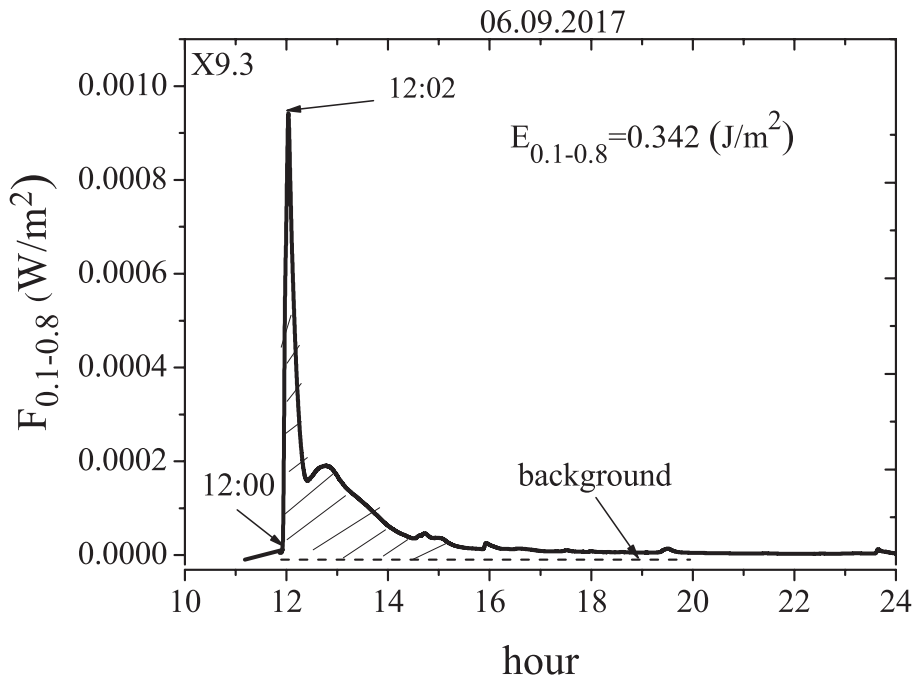}}
 \caption{The flare of Sept. 6, 2017 was the maximum in
terms of x-ray classification in cycle 24, X9.3. The X
axis is Greenwich time GMT.}
{\label{Fi:Fig1}}
\end{figure}

\begin{figure}[tbh!]
\centerline{
\includegraphics[width=120mm]{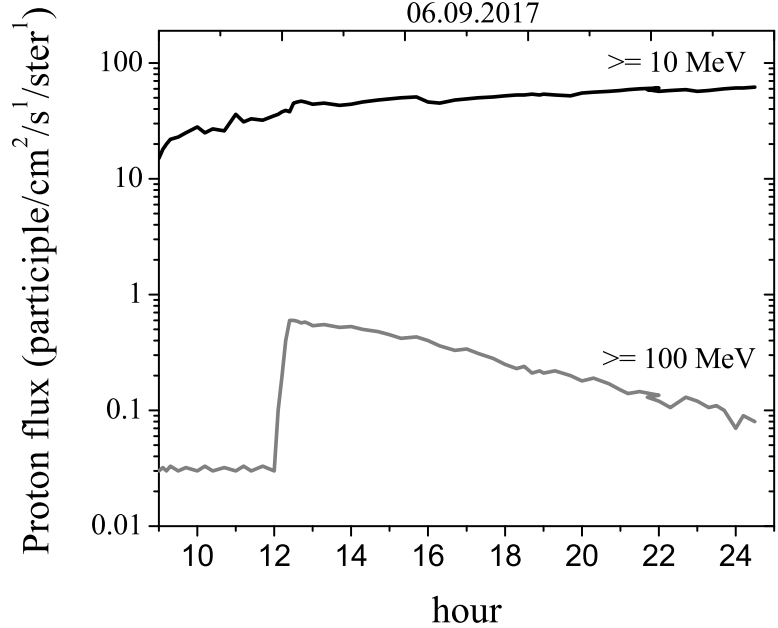}}
 \caption{ The flare of Sept. 6, 2017. (top) Flux of
protons with energies $\geq 10 MeV$; (bottom) flux of
protons with energies $\geq 100 MeV$.}
{\label{Fi:Fig2}}
\end{figure}

\section{The flares in September 2017}

On August 29, group 2673 with an area of 70 MH (1 MH is one millionth of the visible hemisphere of the
sun) emerged from the eastern limb in the southern hemisphere with a total of 1 spot. After two days the area of
this group fell to 60 MH, while the number of spots increased to 4 and the magnetic configuration became more
complicated. On September 3, the group reached the central meridian with an area of 130 MH and 10 spots. Over
the following days, the area of the group increased to 680 MH and the number of spots increased to 28, while the
magnetic configuration became still more complicated. A series of solar flares began on Monday, September 4, and
on that day 7 large class M flares were recorded.

On September 6, this group had 33 spots and an area of 880 MH, and was already in the western hemisphere
near the central meridian. At 12:10 Moscow time, an X2.2 flare with a duration of 20 min was recorded in the group
and at 15:02, another more powerful X9.3 flare lasting 17 min (see Figs. 1 and 2). Both flares were accompanied
by proton events. Over the last twenty years, only five flares with larger amplitudes than X9.3 have been recorded,
and the last of these, of class X17.0, took place almost 12 years ago (September 7, 2005).

Figure 1 is a plot of the radiative flux in the interval 0.1-0.8 nm (linear scale). The total energy arriving from
the flare to the earth, calculated as the area under the shaded time dependence of the flux in the 0.1-0.8 nm range
minus the background level, is indicated in the figure.
Against the background of a prolonged influx of protons from an M5.5 flare, on Sept. 6, 2017 at 12:55 GMT,
protons from an X9.3 flare began to appear over a wide range of energies. For protons with energies $\geq 100 MeV$,
this was a significant event that occurred for the first time since 2014.

Figure 2 illustrates the proton event caused by the flare of Sept. 6, 2017, based on data from observations
of proton fluxes with the GOES-15 satellite at energies $\geq 10 MeV$ and $\geq 100 MeV$. It can be seen that after a
maximum of protons with energies $\geq 10 MeV$ at 12:35 GMT, a gradual decrease in the proton flux began. For
protons with energies $\geq 100 MeV$, the enhanced flux of protons from this flare was superimposed almost unnoticed
against the increased flux of protons from the flare of Sept. 7, after which a further increase in the flux of protons
with energies $\geq 10 MeV$ set in.

On Sept. 10, 2017, at about 16 h Greenwich time, there was yet another very powerful flare of class X8.2 (see
Figs. 3 and 4). It was a continuation of the very powerful X9.3 flare of Sept. 6 and was the next to most powerful
since 2005, being less powerful only than its immediate predecessor of four days before. This flare, which was
associated with the same active region 7623, took place essentially in the limb; on the next day, the active region
2673 went past the edge of the sun, but the hard proton source associated with this flare and located above the active region in the corona, continued to shine above the limb for more than a week. The radiative flux in the 0.1-0.8 nm
range is indicated in Fig. 3 (linear scale). The total energy $E_{0.1-0.8}$ arriving on the earth from the flare is indicated
in the figure; it was calculated as the area under the shaded time dependence of the flux in the 0.1-0.8 nm range.

On Sept. 10, 2017, at roughly 16:30 GMT, proton fluxes from an X8.2 flare began to appear over a wide range
of energies with energies considerably higher than those produced by the X9.3 flare of Sept. 6, 2017.

The enhanced fluxes of protons in the flare of Sept. 10 reached record levels for this cycle, both for protons
with energies $\geq 10 MeV$10 MeV and for the harder $\geq 10 MeV$ 100 MeV protons. The proton fluxes in the flare of Sept. 10 exceeded
those in the flare of Sept. 6 by more than an order of magnitude (see Fig. 4) and this was a record level for all of
cycle 24. The effect of these protons on the earth's atmosphere was grandiose: the magnetic storms following this
flare were also of record magnitude and duration. kp, the index characterizing the degree of perturbation of the earth's
magnetic field reached 6 units over three days according to NOAA/SWPC data (see the archive at the site
http://www.n3kl.org/sun/noaa\_archive).

It is likely that the effect of this proton flare on the earth's atmosphere would have been even greater if the
proton source had not moved together with the flare region beyond the sun's limb.
We emphasize that these powerful flares have taken place against the background of a solar minimum during
cycle 24. The active region 2673 existed for another two revolutions of the sun, but further significant flare activity
in this region was not observed.

\begin{table}
\caption{Parameters of the 21 Large Solar Flares of 2011-2012 Based on Observations
from GOES-15 in the 0.1-0.8 nm Range}
\begin{center}
\begin{tabular}{clclclclclclcl}

\hline

Date of flare/& Background level& Onset of flare& Flare maximum& $E_{0.1-0.8}$ \\
Region of initial& $(W/m^2)$/time & $(W/m^2)$/ time&  $(W/m^2)$/ time& $(J/m^2)$ \\
energy release& & & & \\
\hline
22.09.11/K & 1.05E-6/09:00 & 7.4E-6/10:35 & 1.5E-4/11:00 & 0.756 \\
15.02.11/K &1.12E-6/01:44 &1.3E-6/01:47 &2.3E-4/01:56 & 0.2628 \\
09.08.2011/K &8.0E-7/07:45 & 1.0E-6/08:00 &7.4E-4/08:05 &0.2574 \\
08.03.2011/K, TR& 1.4E-6/18:05 & 2.0E-6/18:10 & 4.5E-5/18:27 & 0.0806 \\
07.03.2011/K & 2.2E-6/19:20 & 3.7E-6/20:12 & 5.0E-5/20:42 & 0.179\\
03.11.2011|/K & 2.06E-6/20:06 & 2.3E-6/20:17 & 2.0E-4|/20:27 &0.168\\
03.08.2011/K & 1.3E-6/17:45 & 2.0E-6/18:10 & 4.4E-5/18:28 & 0.148\\
24.09.2011/K & 1.8E-6/09:31 & 4.0E-6/9:34 & 1.9E-4/09:40 & 0.143\\
06.09.2011/K & 1.8E-6/22:03 & 2.0E-6/22:13 & 2.2E-4/22:20 & 0.118\\
04.08.2011/K & 6,0E-7/03:43 & 2.0E-6/03:44 & 9.5E-5/03:57 & 0.112\\
07.09.2011/K, TR & 6.0E-7/22:13 &3.0E-6/22:35 & 1.8E-4/22:38 & 0.1008\\
08.03.11-1/K & 1.87E-6/10:33 & 3.0E-6/10:30 & 5.4E-5/10:44 & 0.0803\\
09.03.11/K, TR& 2.4E-6/23:00 & 3.0E-6/23:16 & 1.6E-4/23:23 & 0.107\\
13.02.2011/K & 6.5E-7/17:26 &2.0E-6/17:32 & 6.9E-5/17:38 & 0.072\\
25.09.2011/K & 3.0E-6/14:55 & 6.0E-6/15:22 & 3.7E-5/15:43 & 0.064\\
08.09.2011/K & 5.0E-7/15:31 &1.0E-6/15:36 & 6.8E-5/15:46 & 0.059\\
18.02.2011/K & 1.2E-6/09:44 &3.0E-6/10:11 & 7.4E-5/10:12 & 0.053\\
26.12.2011/TR& 1.13E-6/20:00 &1.6E-6/20:15 & 2.4E-5/20:30 & 0.0422\\
24.02.2011/K, TR& 4.0E-7/07:22 & 1.0E-6/07:26 & 3.6E-5/07:35 & 0.0421\\
14.03.2011/K, TR& 5.0E-7/19:32 & 3.0E-6/19:46 &4.4E-5/19:52 & 0.022\\
31.12.2011/K, TR& 6.8E-7/13:09 &1.7E-6/13:12 &2.5E-5/13:15 & 0.0126\\

\hline

\end{tabular}
\end{center}
\end{table}

\section{Large flares in cycle 24. The time of onset and maximum of a flare in the 30.4 nm and 9.4 nm lines and in the interval 0.1-0.8 nm}

As opposed to cycles 21-23, weak flare activity was observed in cycle 24: a total of 133 X-class >M5.0 flares,
of which 49 were X-class >X1. The largest number of these flares was observed near the first and second maxima
of the cycle (of the 10 largest flares, only two took place on the falling branch: these are the largest flares in cycle
24 which occurred on September 6 and September 10, 2017) [5,6]. By comparison with cycles 21-23, the largest
X-class x-ray flares >X15 were observed on the falling branches of cycles 21 and 23, as well as in the maximum of
cycle 22 [7-9]. Observations of flares on satellites with high temporal resolution can be used to study the delay in
the onset time of flares in lines belonging to the lower part of the sun's atmosphere (chromosphere and transition
region) and in the upper part (corona). SDO/EVE observations in extreme ultraviolet lines are available with
averaging over 1 min. GOES-15 observations over 0.1-0.8 nm and 0.05-0.4 nm are available with a time resolution
of 2.5 s. Thus, for each of the flares being studied, we can obtain data on the times of the flare onset and maximum
in the chosen lines.
Figure 5 shows the x-ray fluxes in 0.1-0.8 nm from GEOS-15 and in 5 lines from SDO/EVE for a large class
M8.8 flare that lasted more than 4 h. This is one of the largest flares in terms of total energy ($E_{0.1-0.8} = 0.389 J/m^2$)
in cycle 24. This figure shows the typical times of the inflections corresponding to the onset of a flare and its
maximum (two maxima can be seen in both the 30.4 and 13.3 nm lines). The times of flare onset and the maximum
indicated in Fig. 5 were refined directly from the time series used to construct the plots. It is clear that the typical time profile of a flare in the GOES-15 data for 0.1-0.8 nm corresponds best to the time profile of the FeXX 13.3 nm
coronal line and only slightly less well to the time profile of the FeVIII 9.4 nm coronal line. The maximum
concentrations of FeVIII and FeXX ions are observed in the upper corona for T~107 K and the emission over 0.1-
0.8 nm is formed roughly in this part of the corona [10,11].
For the flare of January 23, 2012, shown in Fig. 5, the corona lines within the 0.1-0.8 nm range (2:12-2:20)
become more intense earlier and the chromospheric lines in the transition region do so later (2:25-2:28).

\begin{figure}[tbh!]
\centerline{
\includegraphics[width=120mm]{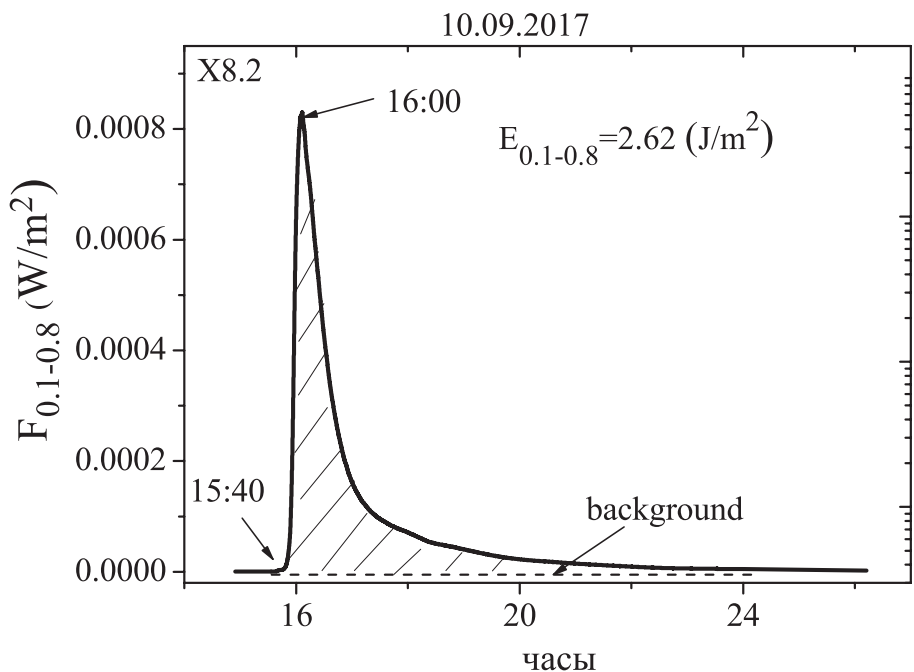}}
 \caption{The flare of Sept. 10, 2017, one of two highest
in terms of x-ray classification in cycle 24. The X
axis is Greenwich time GMT.}
{\label{Fi:Fig3}}
\end{figure}

\begin{figure}[tbh!]
\centerline{
\includegraphics[width=120mm]{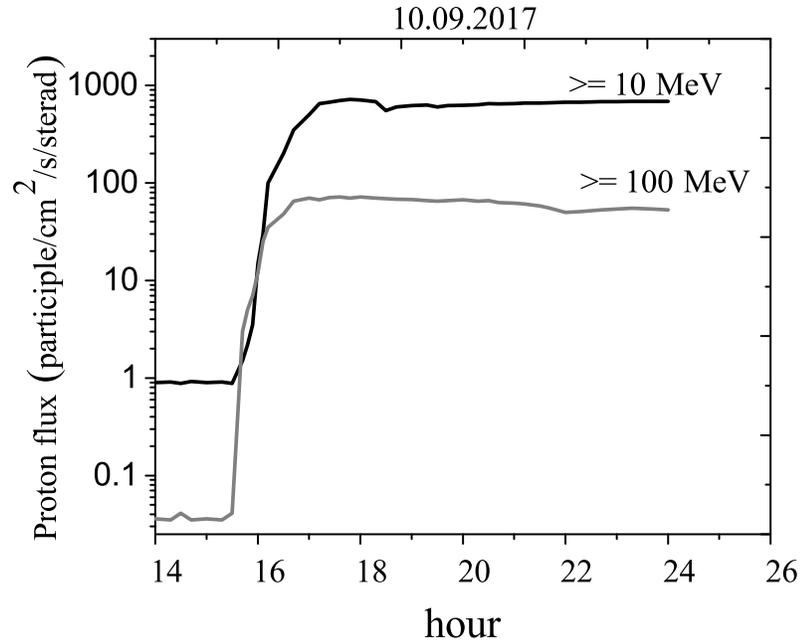}}
 \caption{The flare of Sept. 10, 2017. From the top: the
flux of protons with energies $\geq 10 MeV$; from the
bottom, the flux of protons with energies
$\geq 100 MeV$.}
{\label{Fi:Fig4}}
\end{figure}

\section{Relationship of the total energy radiated by flares in the 0.1-0.8 nm range to the primary energy release region}

For further analysis of the region of primary energy release for the 21 flares we have chosen the lines at 30.4
and 9.4 nm. On one hand, these lines are formed at different altitudes in the sun's atmosphere (30.4 nm is from the transition region and 9.4 nm is a corona line). On the other hand, these lines have additional useful properties: the
30.4 nm line is one of the strongest in the UV and plays a significant role in the formation of the earth's ionosphere,
while the 9.4 nm line is very sensitive to the sun's flare activity and is an order of magnitude more intense than the
13.3 nm line. The series of observations in the 30.4 and 9.4 nm lines was broken in May 2014 because of failure
of some of the measurement instrumentation on the SDO satellite.
We collected data on 21 flares of class >M5 (observations in 2011-2012) and analyzed them in the 30.4 and
9.4 nm lines an the 0.1-0.8 nm range.
14 of these 21 large flares (indicated by the letter K in the first column of Table 1) were similar to the flare
of January 23, 2012, for which the intensification of the fluxes began with the 9.4 nm corona line and the 0.1-0.8 nm range. As an example of such a flare, we consider the X1.9 flare in the corresponding UV lines and x-ray range
which took place on November 3, 2011 (see Fig. 6).

Figure 6 shows the class X1.9 flare of November 3, 2011, which lasted a fairly long time (about one and a
half hours) and had a high total energy incident on the earth in the 0.1-0.8 nm range, $E_{0.1-0.8} = 0.167 J/m^2$. The first
part of Fig. 6 shows the same flare in the 30.4 and 9.4 nm lines. A comparison of the temporal fluxes from the flare
of November 3, 2011 in the 0.1-0.8 nm range and the 30.4 and 9.4 nm lines showed that the initial intensification
of the flare took place in the corona, and after 3-4 min, in the transition region. Here the flare maximum coincides
in time in both lines and in the 0.1-0.8 nm range.
An analysis of 6 large flares out of the 21 studied here (indicated by the symbols K for corona and TR for
transition region in Table 1) showed that for these flares the intensification of the fluxes occurs roughly at the same
time (to within 1 min) in the observed lines and x-ray interval. As an example of these flares, we examine the flare
of March 9, 2011 (see Fig. 7). This flare is not very long (less than an hour) and, therefore, not highly energetic.

A comparison of the time variations in the fluxes from the flare of March 3, 2011, in the 0.1-0.8 nm range
and the 30.4 and 9.4 nm lines showed that the initial intensification of the flare is essentially simultaneous in the
corona and in the transition region. The flare maximum is slightly delayed in the 9.4 nm line.
It can be seen clearly in one of the smallest M3.5 flares that occurred on February 24, 2011, and lasted about
an hour (indicated by the symbol TR in Table 1), that the intensification begins in the lower part of the atmosphere
in the 30.4 nm line in the transition region, and later, after 2-3 min, in the coronal line at 9.4 nm and in the 0.1-
0.8 nm range (see Fig. 8).
Thus, a relationship shows up between the energy of a flare and the region from which it initially releases
energy. It is evident that more data on flares including weaker flares beginning with class M1 will be needed to
confirm the assumption that the initial energy release region for low energy flares is not in the corona, but in the
chromosphere and transition region.

\begin{figure}[tbh!]
\centerline{
\includegraphics[width=110mm]{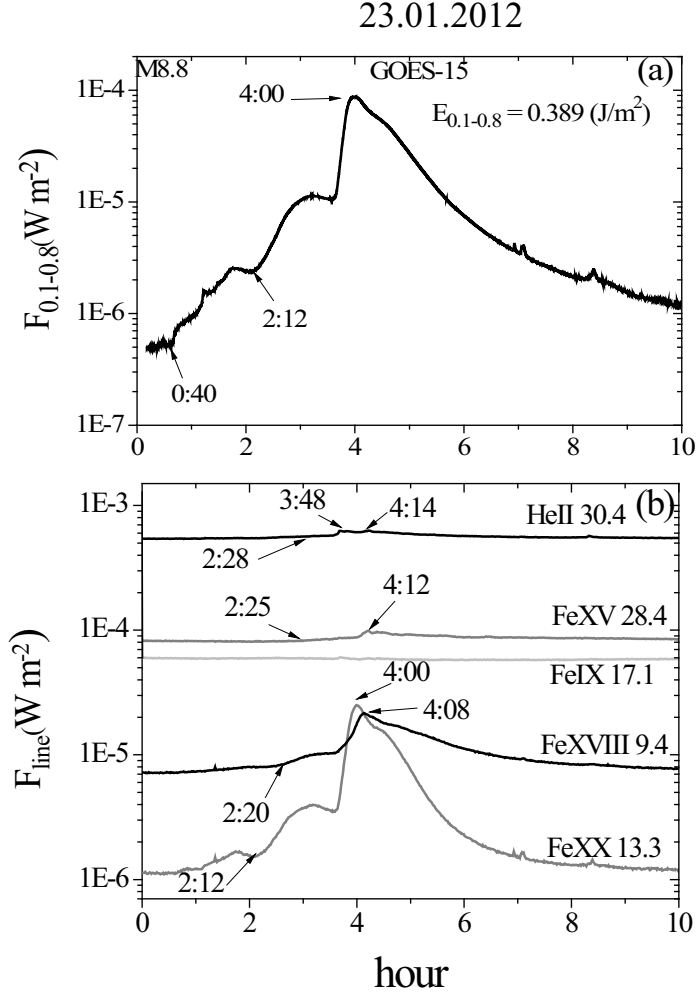}}
 \caption{The M8.8 flare of January 23, 2012. (a) GOES-15 flux
in 0.1-0.8 nm; (b) SDO\/EVE fluxes in UV lines.}
{\label{Fi:Fig8}}
\end{figure}

\begin{figure}[tbh!]
\centerline{
\includegraphics[width=110mm]{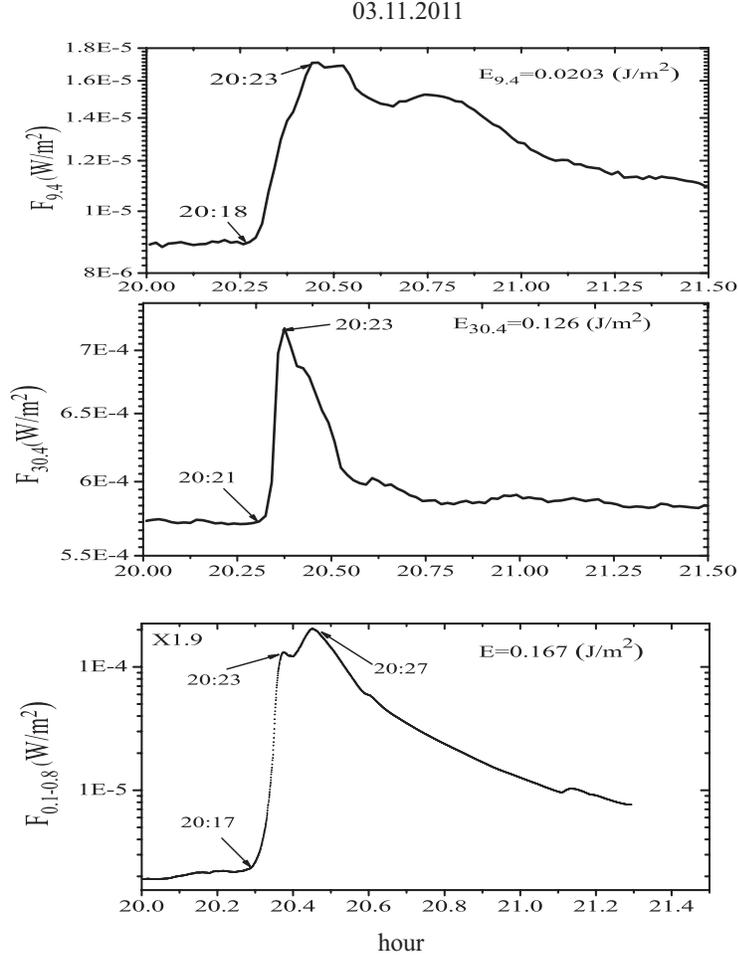}}
 \caption{The X1.9 flare of November 3, 2011.}
{\label{Fi:Fig8}}
\end{figure}

\begin{figure}[tbh!]
\centerline{
\includegraphics[width=110mm]{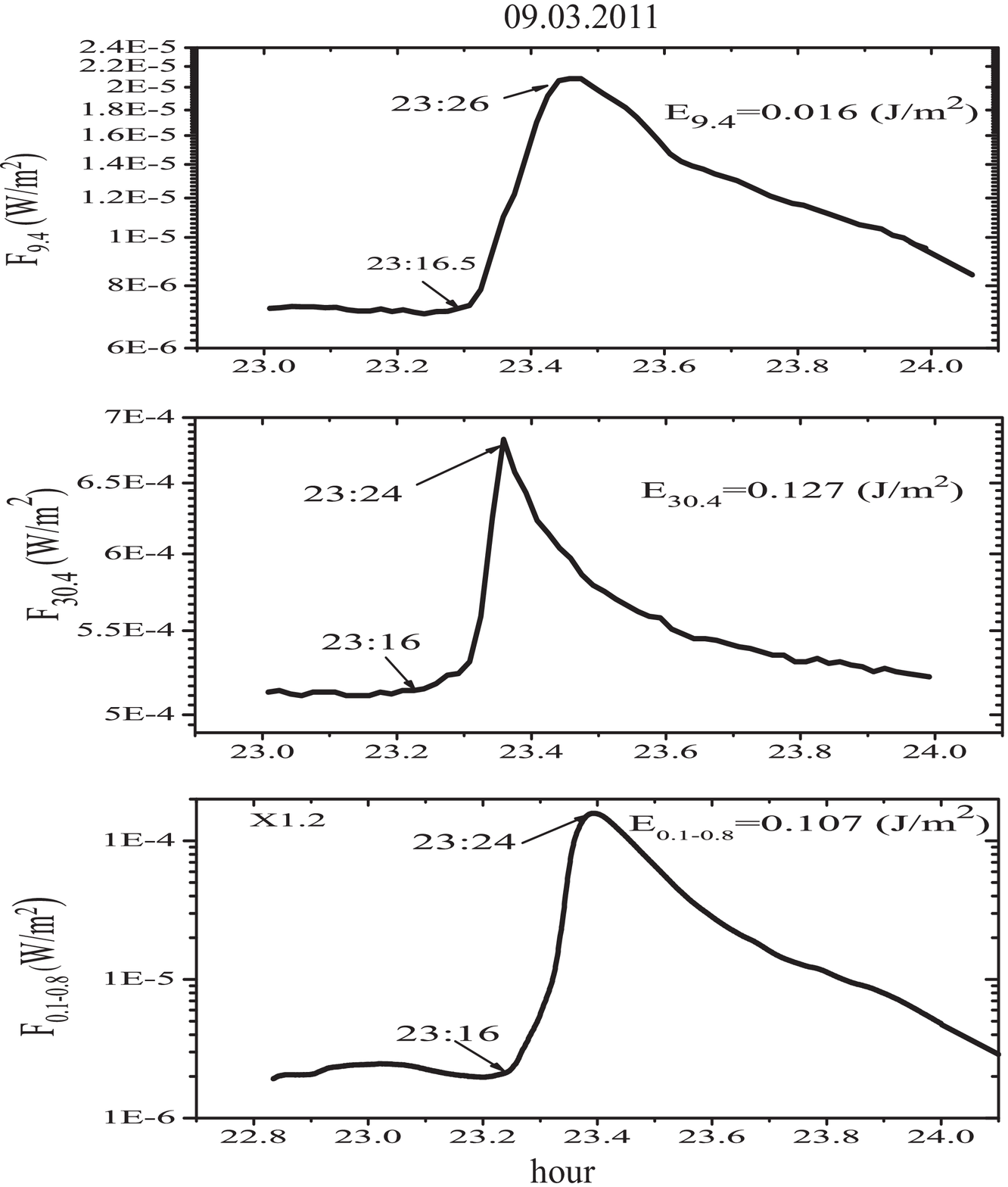}}
 \caption{The X1.2 flare of March 9, 2011.}
{\label{Fi:Fig8}}
\end{figure}

The parameters of the 21 flares that we examine here are listed in Table 1. Data on the background flux level
and on the magnitude and time of the beginning and maximum of the flares in the 0.1-0.8 nm range are listed for
each flare.
The flares in our sample are among the relatively large flares during 2011-2012. We emphasize that in Table
1 they are arranged in order of decreasing total energy emitted in the 0.1-0.8 nm range. The energy $E_{0.1-0.8}$ in that
range was calculated for each flare as the area under the time dependence of the flux from the flare minus the
background flux. For the flares of September 6 and September 10, 2017, these are the shaded areas under the plot
of the total energy $E_{0.1-0.8}$ shown in Figs. 1 and 3 (there were no UV data from SDO/EVE during 2017).
In Table 1, we have identified the flares in terms of the primary energy release region.
The symbol K in column 1 of Table 1 corresponds to flares where the intensification of the flare fluxes began in the corona. The symbols K and TR together correspond to flares where the intensification of the flare fluxes took
place roughly simultaneously in the corona and in the transition region.
The class M3.5 flare of February 24, 2011, with the TR symbol alone corresponds to an earlier intensification
of the flux in the 30.4 nm line of the transition region, followed by the 9.4 nm corona line and the 0.1-0.8 nm interval.
It is clear that the energy $E_{0.1-0.8}$ (fifth column of Table 1) from a flare is associated with the initial phase of
development of the flare process, i.e., the height in the atmosphere where the intensification in the lines begins. Table
1 shows that for the more energetic flares, the intensification begins in the corona. For the less energetic ones, the
flare begins in the corona or simultaneously in the corona and the transition region. For one of the weakest flares
in this sample (No. 19 of the 21), the intensification of the flare flux begins in the lower part of the atmosphere.
A similar dependence on the class of a flare (its peak amplitude, column 4 of Table 1) is not so evident, since powerful
x-ray flares of class X can be of short duration, while longer lasting class M flares can have large values of.

\section{Interrelationship of the energies radiated by flares in the 30.4 and 9.4 nm lines and in the 0.1-0.8 nm range}

Table 2 lists the total energies incident on the earth in the 30.4 and 9.4 nm lines and in the 0.1-0.8 nm range
from 25 large flares. The energies $E_{0.1-0.8}$, $E_{30.4}$, and $E_{9.4}$ are calculated for each flare as the area under the time
dependence of the flux emitted from the flare minus the background flux. Besides the date of the flares, column 1
indicates the flares which were accompanied by ejection of protons (pr).
Table 2 shows that the energy of a flare is related to whether the flare is accompanied by a proton event.
$E_{0.1-0.8}$ is approximately equal to $0.1 J/m^2$ which is a typical value below which flares are not associated with subsequent ejection of protons.
Figure 9 shows the relationship among the energies $E_{0.1-0.8}$, $E_{30.4}$, and $E_{9.4}$. We see a fairly close relationship
between the calculated energies. Evidently, the relationship shown in Fig. 9 can be refined by adding new flares
with large energies (the flares that took place up to mid 2014, when complete observations of the UV lines were made
with SDO/EVE at the same time as the GOES-15 observations). Since these GOES 0.1-0.8 nm observations are
essentially available in real time, while the SDO/EVE UV observations were cut off in mid 2014, the relationship
between the energies is important for estimating the energies emitted in the lines based on $E_{0.1-0.8}$.

\begin{figure}[tbh!]
\centerline{
\includegraphics[width=110mm]{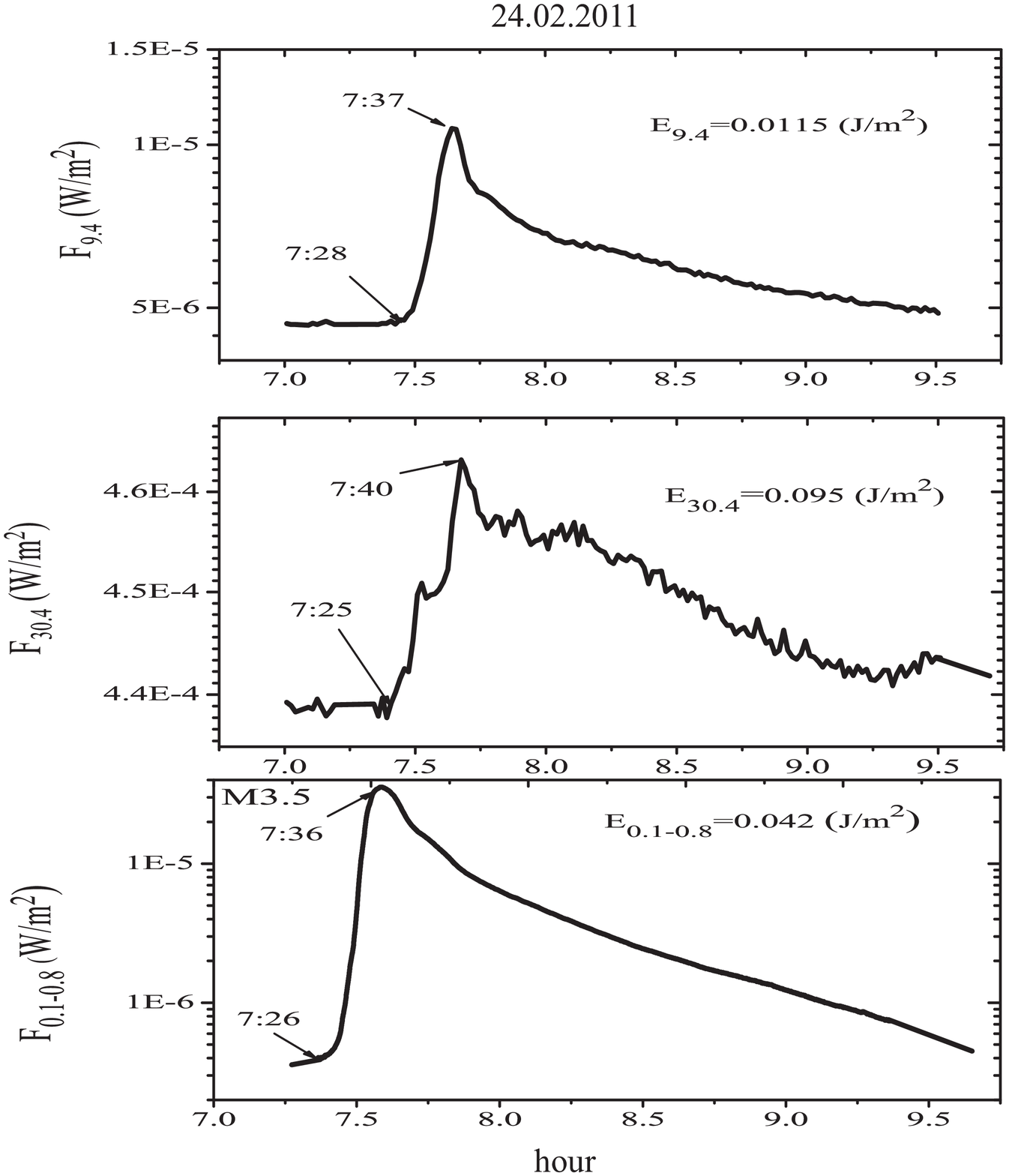}}
 \caption{The M3.5 flare of February 24, 2011.}
{\label{Fi:Fig8}}
\end{figure}

\begin{figure}[tbh!]
\centerline{
\includegraphics[width=110mm]{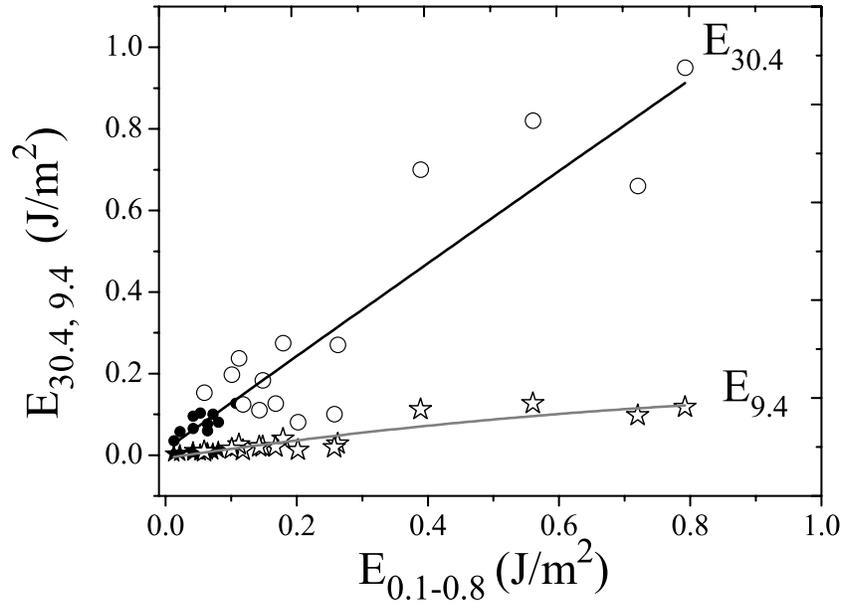}}
 \caption{The relationship between the total energy
radiated by a flare in the UV lines at 30.4 and 9.4
nm and in the 0.1-0.8 nm x-ray range. The circles
correspond to the 30.4 nm line and the stars to the
9.4 nm line. The large hollow circles and stars
indicate that the flares were accompanied by proton
ejection.}
{\label{Fi:Fig9}}
\end{figure}

\begin{table}
\caption{Calculated Energies for 25 Large Flares During 2011-2012
Based on Data from GOES-15 and SDO/EVE}
\begin{center}
\begin{tabular}{clclclclclcl}

\hline
Date of flare&$E_{0.1-0.8}(J/m^2)$ &$E_{30.4}(J/m^2)$& $E_{9.4}(J/m^2)$\\ 
\hline
12.07.2012 pr& ~~0.792 &0.956& 0.118\\
22.09.2011 pr& ~~0.756 &0.66& 0.097\\
23.01.2012 pr& ~~0.389 &0.91& 0.1123\\
15.02.2011 pr& ~~0.263 &0.27& 0.028\\
09.08.2011 pr& ~~0.257 &0.099& 0.0187\\
07.03.2011 pr& ~~0.179 &0.274& 0.024\\
03.11.2011 pr& ~~0.168 &0.126& 0.0203\\
03.08.2011 pr& ~~0.148 &0.183& 0.0205\\
24.09.2011 pr& ~~0.143 &0.11& 0.021\\
31.12.2011 pr& ~~0.126 &0.035& 0.0031\\
23.10.2012 pr& ~~0.119 &0.084& 0.0195\\
06.09.2011 pr& ~~0.118 &0.124& 0.0116\\
04.08.2011 pr& ~~0.112 &0.237& 0.256\\
09.03.2011 pr& ~~0.107 &0.127& 0.016\\
07.09.2011 pr& ~~0.101 &0.197& 0.0154\\
08.03.2011-2 & ~~0.081 &0.012& 0.0179\\
08.03.2011-1 & ~~0.080 &0.081& 0.0124\\
13.02.2011 & ~~0.072 &0.101& 0.082\\
25.09.2011 & ~~0.064 &0.0761& 0.0078\\
08.09.2011 pr & ~~0.059 &0.153& 0.009\\
18.02.2011 & ~~0.053 &0.103& 0.006\\
26.12.2011 & ~~0.042 &0.065& 0.0054\\
24.02.2011 & ~~0.042 &0.095& 0.0115\\
14.03.2011 & ~~0.022 &0.057& 0.0045\\
20.10.2012 & ~~0.064 &0.594& 0.0098\\

\hline

\end{tabular}
\end{center}
\end{table}

\vskip12pt
\section{Conclusions}
\vskip12pt

1. Cycle 24 is characterized by low flare activity. The number of large flares was roughly half that in cycles
22 and 23. The largest flares in September 2017 were of classes X9.4 and X8.3, while class X13-X17 flares occurred
in cycle 23.

2. The most important characteristic of the flares is the total energy from them incident on the earth's surface
per $m^2$ and is the input parameter for analyzing the effect of a flare on the earth's upper atmosphere.

The flare energies in the 30.4 nm transition region line and the 9.4 nm corona region line are closely related
to one another and to the x-ray flux in the 0.1-0.8 nm region, while the amplitudes of the flares in these lines and
the x-ray range are not distinctly related. The energy of a flare determines whether it is a proton flare: beginning
with $E_{0.1-0.8} \geq 0.1 J/m^2$, the flares are accompanied by proton emission.
3. The energy of a flare is related to the region in which initial intensification of the flux of flare emission
takes place (based on an analysis of the 30.4 nm transition region line and the 9.4 nm corona line, as well as the
emission in the 0.1-0.8 nm range): when the flare energy is higher, it is more probable that the region of the initial
intensification is in the corona.

\end{document}